\documentclass[twocolumn]{aastex62}
\usepackage{ulem}
\usepackage{amsmath}

\received{2018 April 24}
\revised{2018 September 7}
\accepted{2018 September 27}
\shorttitle{A limit on gas accretion onto super-Earth cores}
\shortauthors{Ogihara \& Hori}

\begin{document}

\title{A limit on gas accretion onto close-in super-Earth cores from disk accretion}

\correspondingauthor{Masahiro Ogihara}
\email{masahiro.ogihara@nao.ac.jp}

\author[0000-0002-8300-7990]{Masahiro Ogihara}
\affiliation{Division of Theoretical Astronomy, National Astronomical Observatory of Japan,
2-21-1, Osawa, Mitaka,
181-8588 Tokyo, Japan}

\author[0000-0003-4676-0251]{Yasunori Hori}
\affiliation{Division of Optical and Infrared Astronomy, National Astronomical Observatory of Japan}
\affiliation{Astrobiology Center, National Institutes of Natural Sciences,
2-21-1, Osawa, Mitaka,
181-8588 Tokyo, Japan}

\begin{abstract}
The core-accretion model predicts that planetary cores as massive as super-Earths undergo runaway gas accretion to become gas giants. However, the exoplanet census revealed the prevalence of super-Earths close to their host stars, which should have avoided runaway gas accretion. In fact, mass-radius relationships of transiting planets suggest that some close-in super-Earths possess H$_2$/He atmospheres of $\sim$ 0.1--10\,\% by mass. Previous studies indicated that properties of a disk gas such as metallicity and the inflow/outflow cycle of a disk gas around a super-Earth can regulate accumulation of a H$_2$/He atmosphere onto itself. In this paper, we propose a new mechanism that radial mass accretion in a disk can limit the gas accretion onto super-Earth cores. Recent magneto-hydrodynamic simulations found that magnetically driven disk winds can drive a rapid gas flow near the disk surface. Such a rapid gas flow may slip out of a planetary core and regulate gas supply to an accreting gas onto the core. We performed \textit{N}-body simulations for formation of super-Earths with accretion of atmospheres in a viscous accretion disk including effects of wind-driven accretion. We found that even super-Earth cores can avoid triggering runaway gas accretion if the inflow of a disk gas toward the cores is limited by viscous accretion. Our model predicts that super-Earths having H$_2$/He atmosphere of $\sim$ 0.1--10\,wt\% form within $\lesssim 1\,{\rm au}$ of the central star, whereas gas giants are born in the outer region. This mechanism can explain the radial dependence of observed giant planets beyond the solar system.
\end{abstract}


\section{Introduction} \label{sec:intro}

A significant number of super-Earth mass planets have been discovered in close-in orbits.
There are two proposed formation pathways of close-in super-Earths: planetary migration and in-situ formation. The former scenario is that planetary embryos with mass of $\sim 0.1-1~M_\oplus$ form in a region distant from the star and then they undergo type I migration towards the close-in region (e.g., \citealt{cossou_etal14}; \citealt{liu_etal15}; \citealt{izidoro_etal17}). In the latter scenario, newborn embryos in the close-in region ($a \lesssim 1 {\rm ~au}$) can quickly grow to super-Earths before disk dispersal (e.g., \citealt{hansen_murray12}; \citealt{ogihara_etal15a}). In-situ forming super-Earths also undergo type I migration in a disk with a power-law density profile like the minimum-mass solar nebular (MMSN) model \citep{weidenschilling77,hayashi81}, leading to a compact orbital configuration \citep{ogihara_etal15a}. \citet{ogihara_etal18b} found that observed orbital distributions of close-in super-Earths can be reproduced by an in-situ formation model, because type I migration of a planet can be significantly suppressed in a disk with a disk profile (e.g., \citealt{ogihara_etal15b,ogihara_etal15c}) altered by magnetically driven disk winds (e.g., \citealt{suzuki_inutsuka09}; \citealt{suzuki_etal10,suzuki_etal16}).

In addition to orbital configurations, planetary compositions are directly linked to planet-forming environments. Mass-radius relationships of transiting planets suggest that super-Earths may possess atmospheres of $\sim 0.1-10$\,\% by mass (e.g., \citealt{lopez_fortney14}). Recent studies on transmission spectroscopy of transiting super-Earths also revealed the presence of a non-negligible atmosphere; for example, GJ~1214b (e.g., \citealt{kreidberg_etal14}), GJ~3470b (e.g., \citealt{ehrenreich_etal14}), and HD~97658b \citep{knutson_etal14}. Absorption features in the atmospheres of such planets suggested that they have either a cloudy/hazy hydrogen-dominated atmosphere or a hydrogen-poor atmosphere. The budget of a H$_2$/He atmosphere that a super-Earth obtains can be another diagnostic tool of exploring their formation histories. 

The primordial H$_2$/He atmosphere of a planet comes from gas accretion in a protoplanetary disk. Accretion of the surrounding disk gas onto a planet is controlled by atmospheric contraction due to radiative cooling. Mass loading of the accreted disk gas enhances gravitational pull, leading to the atmospheric growth of a planet in a runaway fashion. Eventually, the disk dispersal or a gap-opening in a disk terminates the gas inflow toward the planet. Making a super-Earth having a $\sim 0.1-10$\,wt\% atmosphere requires deceleration and/or suppression of gas inflow into a massive core \citep{ikoma_hori12}. For example, high opacities of dust grains in a disk reduce accumulation of a disk gas by a massive core \citep{lee_etal14,lambrechts_lega17}. Gas dynamics around a planet embedded in a disk would be another adjustment knob for the gas flow. \citet{ormel_etal15} found that an isothermal gas flow from high altitude might escape from the gravitational sphere of a planet before accreting onto it. A rapid recycling of a disk gas regulates gas accretion onto a massive core, whereas in a non-isothermal case, this recycling process would be limited by an entropy gradient between the ambient disk gas and the atmosphere (\citealt{cimerman_etal17}; \citealt{kurokawa_tanigawa18}).

According to recent magneto-hydrodynamic (MHD) simulations, the gas flow structure in a disk have two layers, which is similar to the layered accretion model (e.g., \citealt{gammie96}). There can be a rapid gas flow driven by a disk wind near the disk surface due to the effect of magnetic braking (e.g., \citealt{bai_stone13}; \citealt{suzuki_etal16}), while the accretion flow would be weaker at the midplane of the disk. In fact, global ideal MHD simulations observed a supersonic accretion flow in the upper layer of a disk \citep{zhu_stone17}. A rapid gas flow near the disk surface may put a brake on runaway gas accretion onto a planetary core.

In this paper, we have investigated how a time-dependent layered accretion in a disk affects formation of super-Earths having H$_2$/He atmospheres. We have performed \textit{N}-body simulations for formation of close-in super-Earths, coupled with disk evolution based on MHD simulations and gas accretion onto a planetary core.
Note that hydrogen can be generated from secondary processes such as degassing via oxidation of metallic iron by water \citep{abe_etal00} and H$_2$O photolysis and escape from a planet via stellar XUV irradiations \citep[e.g.][]{lammer_etal13} and interactions between the stellar plasma and a hydrogen atmosphere \citep[e.g.][]{kislyakova_etal13}. However, a long-term evolution of a super-Earth's atmosphere after the formation stage ($\gtrsim 10$\,Myr) is beyond the scope of this paper.

The paper is structured as follows: We present detailed descriptions of models in Section~\ref{sec:model} . Results of gas accretion rates onto planetary cores and their orbital evolution are shown in Section~\ref{sec:n-body}. The radial dependence of the occurrence rate of gas giants and the validity of our model are discussed in Section~\ref{sec:discussion}. We summarize this study in the last section.

\section{Model} \label{sec:model}

\subsection{Disk evolution} \label{sec:disk}
For the evolution of a protoplanetary disk including effects of magnetically driven disk winds, we adopt \citet{suzuki_etal16}'s model in which the diffusion equation is given by
\begin{eqnarray}
\label{eq:diffusion}
\frac{\partial \Sigma_{\rm g}}{\partial t} &=& \frac{1}{r} \frac{\partial}{\partial r} \left[\frac{2}{r\Omega} \left\{ \frac{\partial}{\partial r} (r^2 \Sigma_{\rm g} \overline{\alpha_{r,\phi}} c_{\rm s}^2) + r^2 \overline{\alpha_{\phi,z}} \frac{\Sigma_{\rm g} H \Omega^2}{2 \sqrt{\pi}} \right\} \right] \nonumber \\
 &-& C_{\rm w} \frac{\Sigma_{\rm g} \Omega}{\sqrt{2 \pi}},
\end{eqnarray}
where $\Sigma_{\rm g}, \Omega, c_{\rm s},$ and $H$ are the gas surface density, the Keplerian angular velocity, the sound speed, and the disk scale height, respectively. The first two terms on the right-hand side induce radial mass accretion in a disk, where the mass accretion of the first and the second terms is called the turbulence-driven accretion and the wind-driven accretion, respectively. 
The third term corresponds to mass loss due to disk wind.
Three parameters, $\overline{\alpha_{r,\phi}}, \overline{\alpha_{\phi,z}},$ and $C_{\rm w}$, denote the effective turbulent viscosity \citep{shakura_sunyaev73}, the angular momentum loss due to wind torque, and the mass loss due to disk winds, respectively (see also \citealt{suzuki_etal16}).
We choose values of each parameter based on results of MHD simulations \citep{suzuki_etal10,bai13}.
We consider $\overline{\alpha_{r,\phi}} = 8\times 10^{-3}$ as a fiducial value in an MRI-active disk and
$\overline{\alpha_{\phi,z}} = 10^{-5} (\Sigma_{\rm g} / \Sigma_{\rm g,ini})^{-0.66}$, where $\Sigma_{\rm g,ini}$ represents the initial gas surface density.
An MRI-inactive case with a smaller $\overline{\alpha_{r,\phi}} (= 8\times 10^{-5})$ is also examined in Appendix~\ref{sec:app1}.
For mass loss rates via a disk wind, we consider upper limits of $C_{\rm w}$ to be $2 \times 10^{-5}$ (MRI-active case) and $1 \times 10^{-5}$ (MRI-inactive case).

%Fig.1
\begin{figure}[ht!]
\plotone{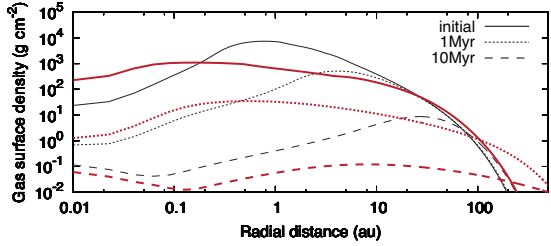}
\caption{Evolution of gas surface density. Red and black lines represent the gas surface density at $t = 0$ (solid), $t = 1$\,Myr (dotted), and $t = 10$\,Myr (dashed) in an MRI-active and an MRI-inactive (see Appendix~\ref{sec:app1}) disk, respectively.}
\label{fig:r_sigma}
\end{figure}

\citet{suzuki_etal16} considered the disk evolution from the collapse stage of a molecular cloud to star formation. They assumed that the initial disk profile, $\Sigma_{\rm g,ini}$, was ten times as massive as the MMSN model. We focus on processes of gas accretion onto planetary cores, which correspond to the middle/late stage of disk evolution. Thus, we use the disk profile at $t = 0.1 {\rm ~Myr}$ given in \citet{suzuki_etal16} for the initial conditions as well as  
our previous studies \citep[e.g.][]{ogihara_etal17}.
Figure~\ref{fig:r_sigma} shows the time evolution of gas surface density in this study. We note that the disk profile is significantly different from a power-law distribution of the MMSN model, specifically that the surface density of a disk gas within $r \lesssim 1\,{\rm au}$ has a flat profile for MRI-active disks.

\subsection{Gas accretion onto a planetary core}\label{sec:gas-accretion}

We investigate how much H$_2$/He gas is accreted by super-Earth cores from the surrounding disk that viscously evolves with magnetically driven disk winds. We adopt semi-empirical prescriptions for gas accretion based on \citet{ogihara_etal15a}.
\citet{ogihara_etal15a} derived the gas accretion rate onto the core from
one-dimensional quasi-static evolution models of a planetary atmosphere, 
which is given by
\begin{eqnarray}\label{eq:mdot_env}
\dot{M}_{\rm env, OMG} = M_{\rm env} \left[\frac{k_3}{3} \frac{1}{t^{1/3}_{\rm run} t^{2/3} + k_3 t} + \frac{k_1}{t_{\rm run}}\right],
\end{eqnarray}
where $M_{\rm env}, M_{\rm core}$, and $t_{\rm run}$ are the atmospheric mass, the core mass of a planetary embryo, and the time when $M_{\rm env}$ reaches the crossover mass (i.e., $M_{\rm env} \sim M_{\rm core}$). The crossover time is given by 
\begin{equation}\label{eq:trun}
t_{\rm run} = T_{\rm run} \left( \frac{M_{\rm core}}{5~M_\oplus} \right)^{-3} {\rm ~yr},
\end{equation}
where $ T_{\rm run}$ is between $10^6$ and $10^7$.
This is a single free parameter in this model, which mainly indicates the effect of grain opacities on atmospheric contraction. The choice of coefficients, $k_1 = (M_{\rm core} / 15~M_\oplus)^{-1}$ and $k_3 = 9$, reproduces results of \citet{ikoma_hori12}, \citet{piso_youdin14}, and \citet{lee_etal14}\footnote{Equation (\ref{eq:mdot_env}) can reproduce the mass accretion rates of \citet{piso_youdin14} and \citet{ikoma_hori12} by using $T_{\rm run} = 8 \times 10^6$ and $T_{\rm run} = 2.5 \times 10^6$, respectively. We can also fit the result of \citet{lee_etal14} by adopting $T_{\rm run} = 2 \times 10^7 {\rm ~yr}$ in Eq.~(\ref{eq:trun}).}.

This gas accretion model does not explicitly include effects of other factors (e.g., disk density) on the gas accretion rate. In practice, the critical core mass and hence the gas accretion rate onto the core depend on gas density and temperature in the ambient disk. We combine these effects into the single parameter $t_{\rm run}$ and make the model simple. Note that the critical core mass is not sensitive to the boundary condition (e.g., disk density); therefore $t_{\rm run}$ mainly expresses the uncertainty in the opacity as stated above. In addition, if the accretion rate of planetesimals onto the core is high, the growth of envelope also depends on the planetesimal accretion. In this study, we consider that the accretion of planetesimals stalls during the late stage of planet formation. This means that there is no extra heating except for gravitational contraction of the envelope of a planet. Combined with the low disk density in the later stage of disk evolution, the outer envelope is likely to be radiative. In this sense, the assumption that the critical core mass (and hence the gas accretion onto cores) is insensitive to boundary conditions such as gas density and temperature is justified.

\subsection{Disk gas flow} \label{sec:flow}

The accretion rate given in Eq.~(\ref{eq:mdot_env}) assumes that the disk can supply all the gas that a core is able to accrete. In fact, the inflow of a disk gas toward the core should be limited by the disk accretion due to turbulence-driven mass accretion in the protoplanetary disk. The accretion rate was taken to be the minimum of $\dot{M}_{\rm env, OMG}$ (Eq.~\ref{eq:mdot_env}) and $\dot{M}_{\rm turb} (\simeq 3 \pi \nu \Sigma_{\rm g})$ in \citet{ogihara_etal15a}, where $\nu$ represents the turbulent viscosity. Thus, the rate of gas accretion onto cores is reduced as the gas density decreases.

%Fig.2
\begin{figure}[ht!]
\plotone{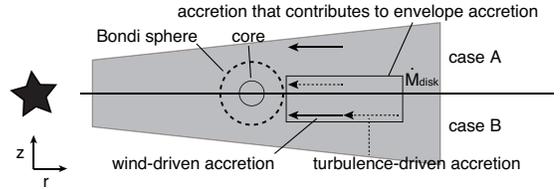}
\caption{Schematic picture of disk accretion patterns in two cases. In case\,A, only the turbulence-drive accretion around a small planet supplies a disk gas to the Bondi sphere of the planet. In case\,B, both the turbulence-driven accretion and the wind-driven accretion flow into the region around the planet.}
\label{fig:gas_flow}
\end{figure}

In this paper, we consider the wind-driven accretion due to the effect of magnetic braking. The upper limit on the radial mass accretion in a disk would be the sum of the turbulence-driven accretion and wind-driven accretion (see case\,B in Figure~\ref{fig:gas_flow}). However, the mass accretion at the disk midplane may be independent of wind-driven accretion that occurs near the disk surface if the vertical structure of gas flow in a disk is considered (case\,A in Figure~\ref{fig:gas_flow}). In case A, the gas flow in the upper layer of a disk moves rapidly because the wind torque is entirely exerted there. In fact, the radial velocity of the wind-driven accretion flow is close to the sound speed \citep{suzuki_etal16,ogihara_etal17,zhu_stone17}, which is comparable to the shear velocity at the Bondi radius. Such a rapid gas flow likely slips out of the core and regulates gas supply to the gas accretion onto it.
Therefore, we consider two cases of disk accretion for the limit of atmospheric accretion as follows. 
\begin{eqnarray}
\label{eq:mdot_disk}
\dot{M}_{\rm disk} = 
\begin{cases}
 \dot{M}_{\rm turb} &{\rm (for~case\,A)}\\
 \dot{M}_{\rm turb} +  \dot{M}_{\rm wind} &{\rm (for~case\,B)}.
\end{cases}
\end{eqnarray}  
The net gas accretion rate on a core $\dot{M}_{\rm env}$ is given by
\begin{equation}
\label{eq:mdot_min}
\dot{M}_{\rm env} = \min (\dot{M}_{\rm env, OMG}, \dot{M}_{\rm disk})
\end{equation}
where $\dot{M}_{\rm env, OMG}$ and $\dot{M}_{\rm disk}$ are given by Eqs.~(\ref{eq:mdot_env}) and (\ref{eq:mdot_disk}).
Case\,A is considered as a lower limit on the mass accretion rate. Note that we purposely use a simplified model and the reality should be more complicated. A discussion in this regard is given in Section~\ref{sec:validity}.

\begin{figure}[ht!]
\plotone{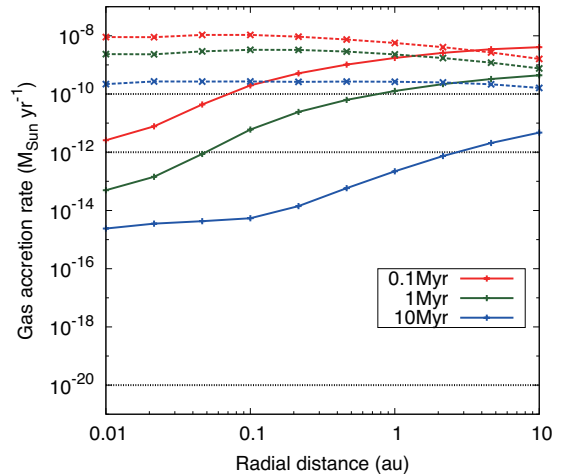}%{mdot_inactive}
\caption{Time evolution of accretion rates. Solid and dashed lines represent mass accretion rates driven by turbulence ($\dot{M}_{\rm turb}$) and by a disk wind ($\dot{M}_{\rm wind}$). Horizontal dotted lines are typical gas accretion rates onto a planetary core with mass of 1, 5, 10\,$M_\oplus$ at $t = 0.1 {\rm ~Myr}$.}
\label{fig:mdot}
\end{figure}

The radial mass accretion rates in the disk due to turbulence-driven accretion and wind-driven accretion are given by (see also Eqs.~(33) and (34) of \citealt{suzuki_etal16})
\begin{eqnarray}
\label{eq:mdot_turb}
\dot{M}_{\rm turb} = - \frac{2 \pi}{\Omega} \overline{\alpha_{r,\phi}} c^2_{\rm s} \Sigma_{\rm g},\\
\label{eq:mdot_wind}
\dot{M}_{\rm wind} = - 2 \sqrt{2 \pi} \overline{\alpha_{\phi,z}} r c_{\rm s} \Sigma_{\rm g}.
\end{eqnarray}
Figure~\ref{fig:mdot} shows the time evolution of the mass accretion rate in an MRI-active disk with $\overline{\alpha_{r,\phi}} = 8\times 10^{-3}$. Since the mass accretion rate in a disk is proportional to a local gas surface density (see Eq.~(\ref{eq:diffusion})), $\dot{M}_{\rm turb}$ becomes low in the close-in region ($r \lesssim 1\,{\rm au}$), as seen in Figure \ref{fig:r_sigma}; for example, $\dot{M}_{\rm turb} < 10^{-8}~M_\oplus ~{\rm yr}^{-1} \sim 10^{-13}~M_\odot ~{\rm yr}^{-1}$ at $r = 0.01 {\rm ~au}$ and $t = 1{\rm ~Myr}$. This is inconsistent with the inferred mass accretion rate onto the star ($\dot{M}_{\rm disk} \sim 10^{-9} M_\odot~{\rm yr}^{-1}$; \citealt{hartmann_etal98}; \citealt{manara_etal16}). On the other hand, the wind-driven accretion is almost independent of radial distance because the angular momentum loss due to wind torque, i.e., $\overline{\alpha_{\phi,z}}$, increases with decreasing $r$ (see Figure 7 in \citealt{suzuki_etal16}).
As a result, the wind-driven accretion overwhelms the turbulence-driven accretion in the close-in region, as shown in \citet{suzuki_etal16}, which means that it plays an important role in determining the disk evolution\footnote{The wind-driven accretion tends to dominate over the mass loss term, namely, the third term on the right-hand side of Eq.~(\ref{eq:diffusion}). We note that in \citet{suzuki_etal16}, the strength of wind mass loss $C_{\rm w}$ is weakened by a factor of about five from results of local MHD simulation without non-ideal MHD effects \citep{suzuki_inutsuka09}. This is because non-ideal MHD effects would reduce the mass-loss efficiency \citep{bai13}. In addition, the parameter $C_{\rm w}$ in the close-in region ($r \lesssim 1~{\rm au}$) can be further constrained by energetics (see Section 2.3 of \citealt{suzuki_etal16}).}.

%Fig.3

%Fig.4
\begin{figure*}[ht!]
\plottwo{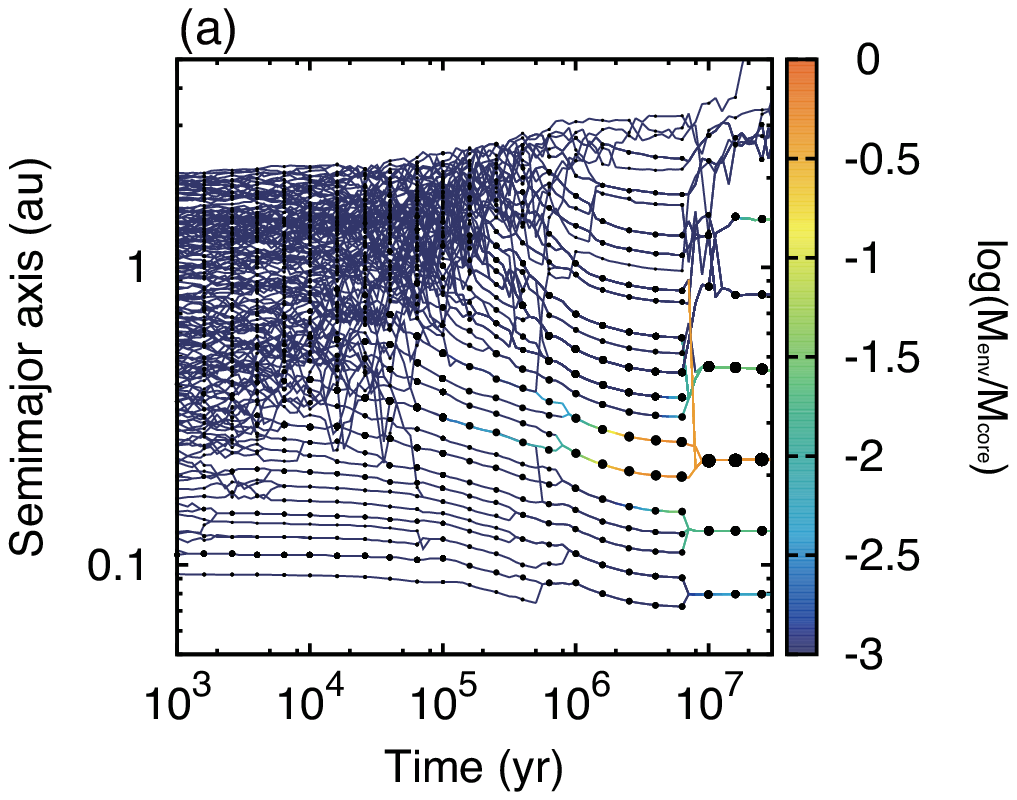}{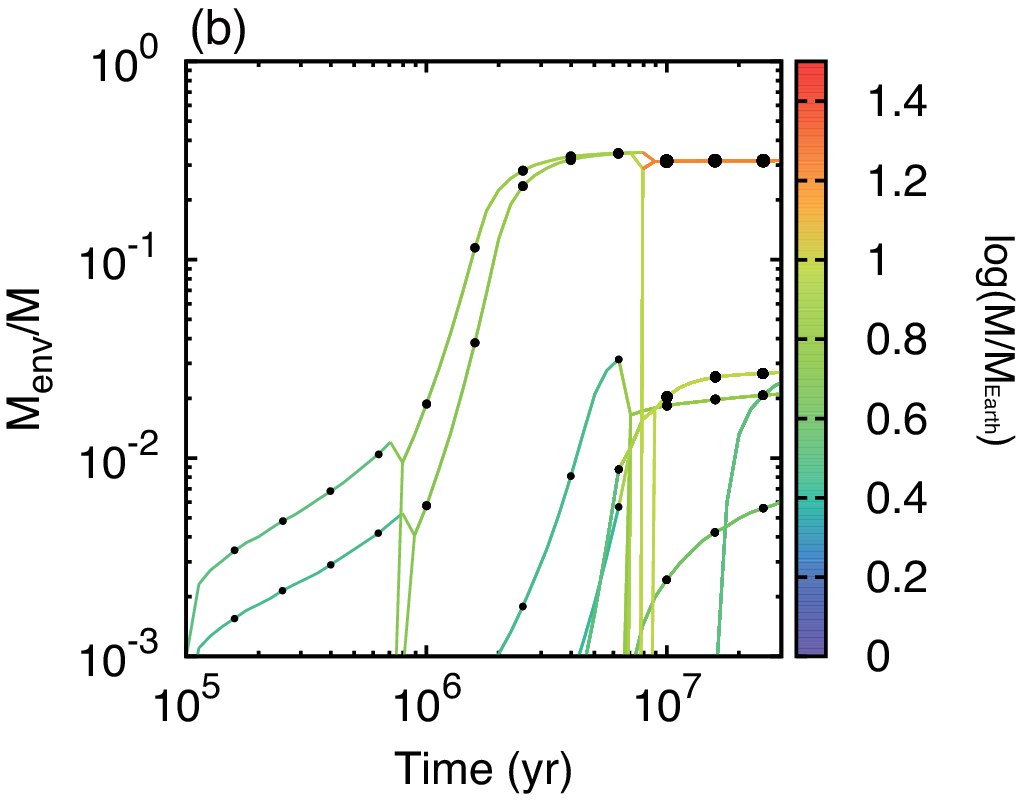}
\caption{Typical results of an \textit{N}-body simulation for case\,A and $T_{\rm run}=10^6$. The left and right panels show the time evolution of semi-major axis and the envelope mass fraction of a planet, respectively.}
\label{fig:run6}
\end{figure*}

Here we estimate the rate of gas accretion onto a core, $\dot{M}_{\rm env, OMG}$, which would help us interpret results of \textit{N}-body simulations in the next section. By substituting $T_{\rm run}=10^6$ into Eq.~(9) of \citet{ogihara_etal15a}, we can obtain $M_{\rm env}$. Combined with Eq.~(\ref{eq:mdot_env}), the maximum envelope accretion rate at $t = 0.1~{\rm Myr}$ is roughly given by $\dot{M}_{\rm env, OMG} \sim 10^{-20}$, $10^{-12}$, $10^{-10}\,M_\odot~{\rm yr}^{-1}$ for $M_{\rm core} =$1, 5, 10\,$M_\oplus$, respectively.
Figure~\ref{fig:mdot} shows that the estimated $\dot{M}_{\rm env, OMG}$ for a core with $M = 1~M_\oplus$ is lower than $\dot{M}_{\rm disk}$. However, in case\,A in which only the turbulence-driven accretion contributes to the gas accretion onto the core, the estimated $\dot{M}_{\rm env, OMG}$ for $M = 5~M_\oplus$ and $10~M_\oplus$ can overwhelm $\dot{M}_{\rm turb}$, especially in the close-in region and during the late stage of disk evolution ($t > 1~{\rm Myr}$). Therefore, the gas accretion onto the core is likely limited by the turbulence-driven accretion in a disk for case\,A. This also happens in an MRI-inactive disk (see Appendix~\ref{sec:app1}). In case\,B in which the wind-driven accretion well as the turbulence-driven accretion affects gas accretion onto a core, $\dot{M}_{\rm disk}$ is expected to be higher than $\dot{M}_{\rm env, OMG}$ by several orders of magnitude in the whole region of a disk.

\section{Results: \textit{N}-body simulations} \label{sec:n-body}

We performed a series of \textit{N}-body simulations of formation of super-Earths around an $1\,M_\odot$ star, coupled with gas accretion onto a planetary core. Numerical settings are the same as those used in \citet{ogihara_etal18b}, in which the gas accretion onto cores was not considered. We initially placed up to 400 planetary embryos with mass of $0.2~M_\oplus$ between 0.1 and 2\,au in an MRI-active disk (see also the Appendix~\ref{sec:app1} for results in the case of MRI-inactive disks).
The total mass of solid material was assumed to be $40~M_\oplus$. \citet{ogihara_etal18b} found that this number of building blocks roughly reproduces observed distributions of close-in super-Earths.
Simulations were performed for both cases\,A and B. In addition, we changed the crossover time $T_{\rm run}$ in Eq.~(\ref{eq:trun}) which  depends on several physical properties (e.g., opacity, disk temperature).
During the early phase of planet formation, the mass of a planetary embryo is not large enough to acquire a dense atmosphere. Thus, we initiated gas accretion onto a core after  $t = 0.1$\,Myr in the same way as \citet{ogihara_etal15a}. The initial amount of atmosphere at $t = 0.1 {\rm ~Myr}$ is calculated by substituting $t=0$ into Eq.~(\ref{eq:mdot_env}). In this study, once a planetary core satisfies the thermal criterion for gap-opening, i.e., $r_{\rm H} \gtrsim H$ \citep[e.g.][]{lin_papaloizou93}, gas accretion onto itself is quenched and type I migration stops. Then the migration regime would change to type II migration. However, the type II migration rate in a thin disk for the late stage of disk evolution is uncertain; therefore, for simplicity, we shut off the migration when the thermal gap opening criterion is satisfied.
We neglected atmospheric loss of a planet via giant impacts during the gas depletion phase, which means that the final mass of a planetary atmosphere in this study provides an upper limit to the amount of H$_2$/He gas that the planet can acquire.

\subsection{Case A: Gas accretion is limited by turbulence-driven accretion}\label{sec:caseA}

Figure~\ref{fig:run6} shows the time evolution of the semi-major axis and the mass fraction of a planet's envelope for our fiducial run ($T_{\rm run} = 10^6$) of case\,A. The color indicates the envelope-to-core mass ratio in Figure~\ref{fig:run6}a and the planet mass in Figure~\ref{fig:run6}b. As shown in \citet{ogihara_etal18b}, planets do not undergo significant type I migration because a positive corotation torque acted on a planet compensates for a negative Lindblad torque in a disk with a flat/positive slope of gas surface density (see Fig.~\ref{fig:r_sigma}) in the close-in region ($r \lesssim 1 {\rm ~au}$). The positive corotation torque exerted on planets with $M \sim$ 1--10\,$M_\oplus$ does not become saturated in an MRI-active disk (see \citealt{ogihara_etal18b} for more detail). Figure~\ref{fig:run6}b shows that after planets accreted some amount of atmospheres, gas accretion onto cores almost ceased. Beyond $t \gtrsim 3 \times 10^6 {\rm ~yr}$, $\dot{M}_{\rm turb} (=\dot{M}_{\rm disk})$ becomes smaller than $\dot{M}_{\rm env, OMG}$ (see Figure~\ref{fig:mdot}) and the accretion of atmospheres is limited by disk accretion. Thus, super-Earths possess H$_2$/He atmospheres of 0.6--30\% by mass in the end. 

%Fig.5
\begin{figure}[ht!]
\plotone{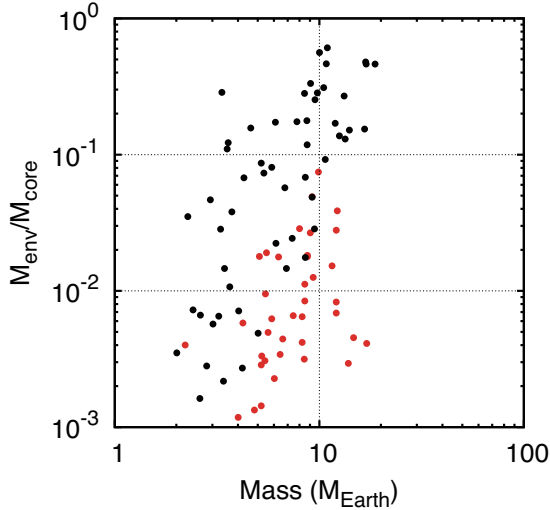}
\caption{Final envelope-to-core mass ratio of a planet for case\,A. Black and red dots represent 10 simulation results for $T_{\rm run}=10^6$ and
$T_{\rm run}=10^7$.}
\label{fig:m_menv}
\end{figure}

We also performed ten runs of simulations each for $T_{\rm run} = 10^6$ and $10^7$. Figure~\ref{fig:m_menv} shows the final envelope mass fraction as a function of planetary mass. A rapid contraction of a planet's atmosphere, namely $T_{\rm run} = 10^6$, results in formation of super-Earths with more massive H$_2$/He atmospheres. However, even massive cores with $M > 10~M_\oplus$ avoid runaway gas accretion. Thus, all the planets retain 0.1--40\% of H$_2$/He envelopes by mass.

\begin{figure}[ht!]
\plotone{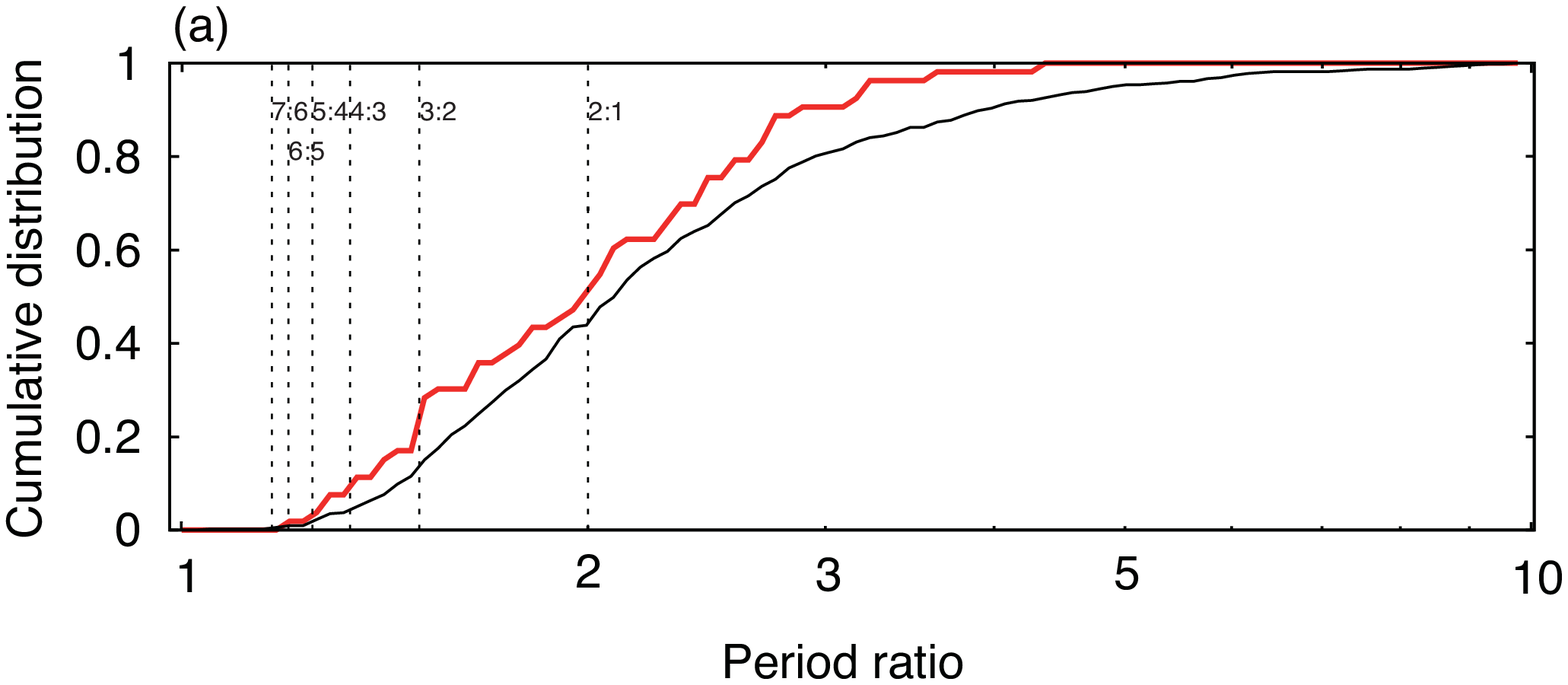}
\plotone{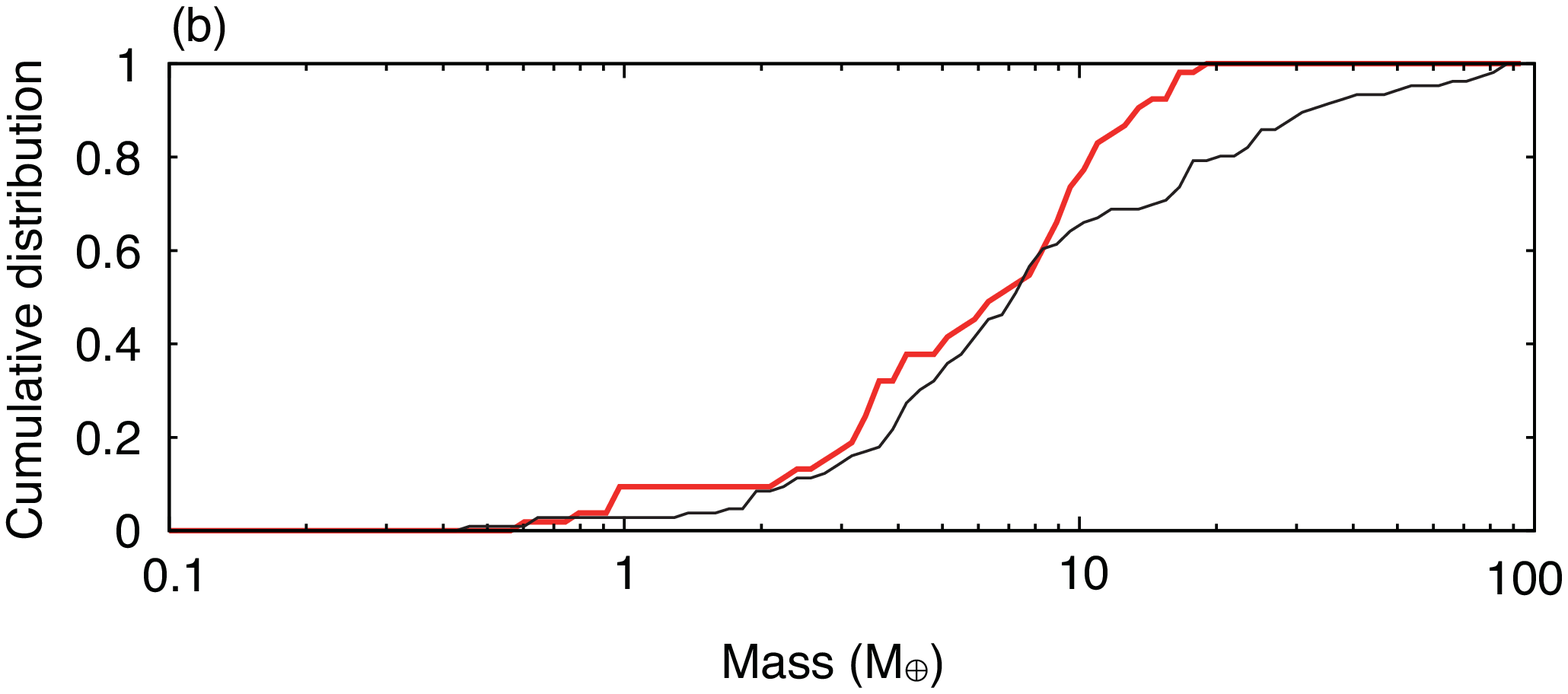}
\caption{Cumulative distributions of planet populations. Panel~(a) shows the period ratio ($P_{\rm out}/P_{\rm in}$) of an adjacent pair of planets. Vertical dashed lines indicate locations of first-order mean motion resonances. Panel~(b) shows the mass distribution of planets. The red and black lines indicate results ($T_{\rm run}=10^6$) for case\,A and observed distributions of close-in planets, respectively.
}
\label{fig:cumulative}
\end{figure}

Figure~\ref{fig:cumulative}a indicates the cumulative distribution of period ratio of adjacent planets $P_{\rm out}/P_{\rm in}$, in which results of ten runs are summarized. Figure~\ref{fig:cumulative}b shows the cumulative distribution of  final mass. We also plotted observed distributions of 1087 confirmed planets in multiple close-in super-Earth systems (data from the NASA Exoplanet Archive). Our results yield planet distributions similar to those of observed close-in super-Earths. The cumulative distributions of the period ratio and the mass are slightly smaller than those in \citet{ogihara_etal18b} (see also Fig.~7 of \citealt{ogihara_etal18b}). This means that the inclusion of the envelope accretion does not trigger the late orbital instability, which leads to additional collisional events between planets in the resonant chain. Note that we examined the growth and orbital evolution of planets until $t= 30~{\rm Myr}$ in our simulations. Subsequent collision events are expected to occur after $t= 30~{\rm Myr}$, increasing period ratios of planets.
\citet{ogihara_etal18b} performed \textit{N}-body simulations for formation of super-Earths until $t= 100~{\rm Myr}$ using the same parameters but without gas accretion onto a planet, and found that the period-ratio distribution is slightly more separated than our current results indicate.
In any case, the final orbital configuration of our simulated planets is consistent with the observed distribution of close-in super-Earths.

%Fig.6

\subsection{Case B: Gas accretion is limited by turbulence-driven and wind-driven accretion}

\begin{figure*}[ht!]
\plottwo{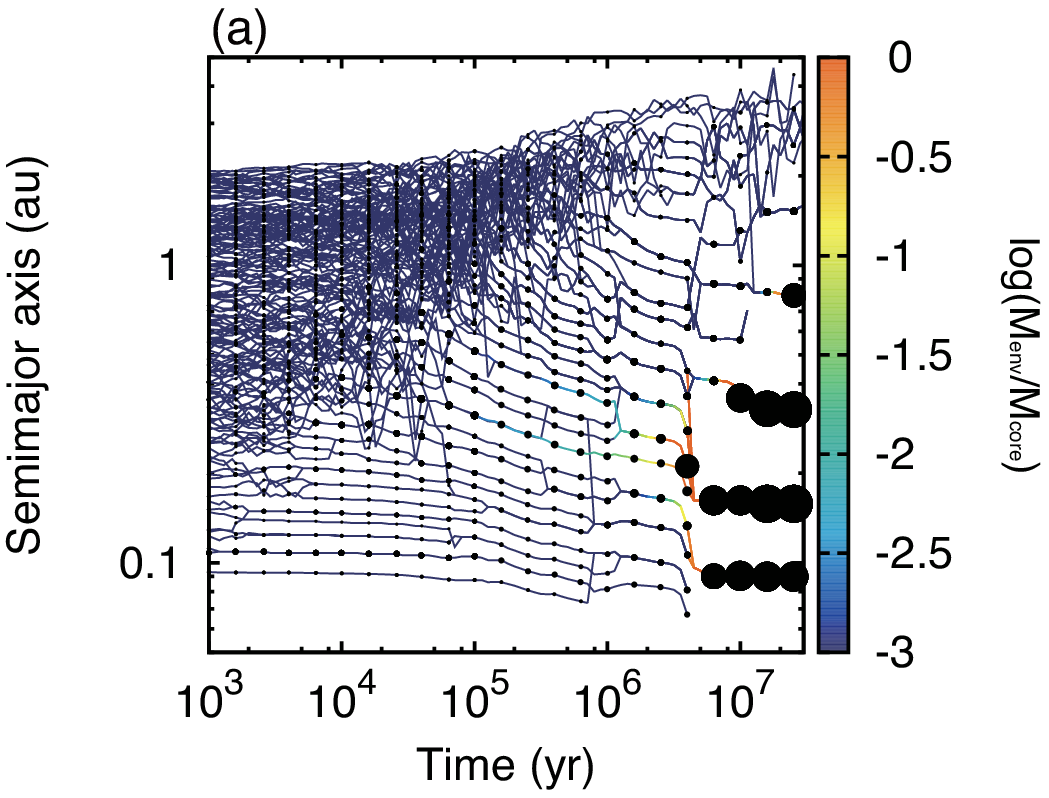}{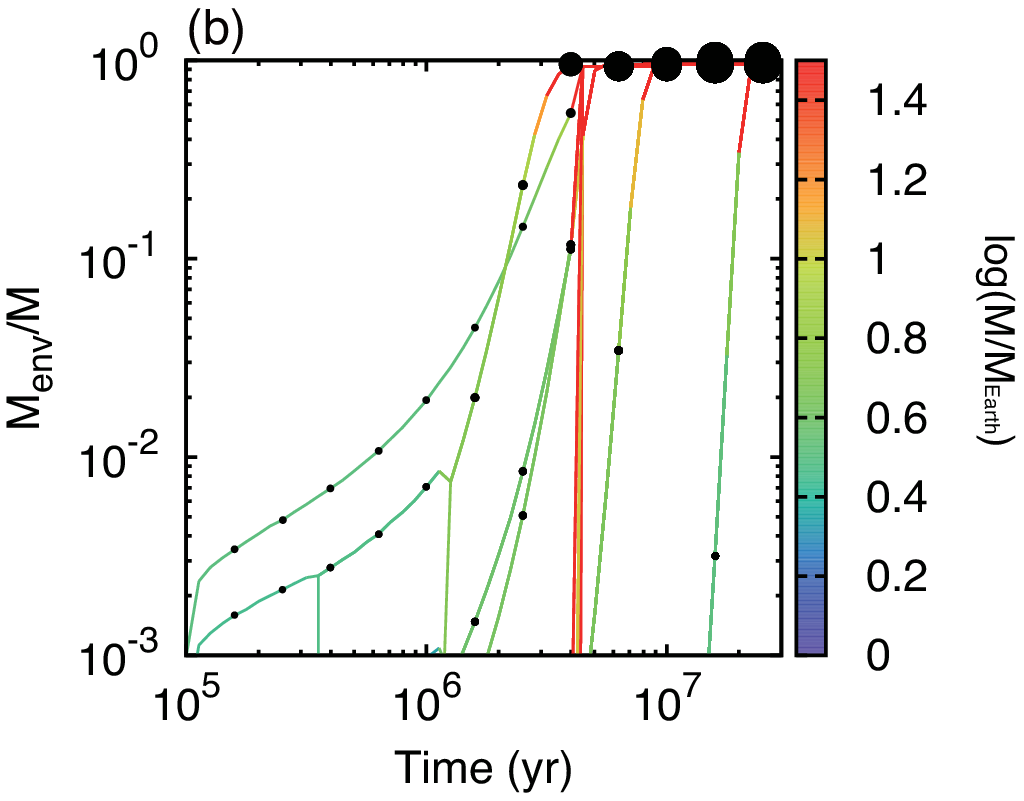}
\caption{Same as Fig.~\ref{fig:run6} but for case\,B.}
\label{fig:run120}
\end{figure*}

Figure~\ref{fig:run120} shows simulation results of case\,B. As expected, planets accreted a significant amount of H$_2$/He gas
because $\dot{M}_{\rm disk}$ is higher than $\dot{M}_{\rm env, OMG}$, as shown in Figure~\ref{fig:mdot}.
At $t \simeq 4~{\rm Myr}$, the envelope masses of massive planets become comparable to their core masses, leading to runaway gas accretion.
As a result, no planets with 0.1--10\% of H$_2$/He atmospheres formed, while four gas giants formed within $1 {\rm ~au}$ from the central star. Although hot Jupiters that reside within 0.1~au may lose their envelopes by mass loss due to a stellar XUV irradiation (e.g., \citealt{owen_wu13}; \citealt{kurokawa_nakamoto14}),
there is a lack of multiple gas giants near a central star, according to planet searches (e.g., \citealt{latham_etal11,steffen_etal12}). 
Therefore, results for case\,B, in which multiple super-Earths undergo runaway gas accretion to be hot Jupiters, are incompatible with observations.

%Fig.7

\section{Discussion} \label{sec:discussion}

\subsection{Inner super-Earths and outer gas giants}

The exoplanet census suggests the prevalence of super-Earths and the rarity of giant giants in the close-in region ($r \lesssim 1 {\rm ~au}$) \citep[e.g.][]{fressin_etal13}. On the other hand, the occurrence rate of giant planets increases with increasing orbital separations from a host star (e.g., \citealt{santerne_etal16}). To explain the observed planet population, several ideas that avoid runaway gas accretion onto super-Earths in the inner region of a disk have been proposed; for example, a negative [Fe/H] gradient in a disk \citep{lee_etal14} and a density gap in a low-viscosity disk \citep[e.g.][]{fung_lee18}.

In this paper, we propose a new mechanism that explains a radial dependence of the observed giant planet population.
If wind-driven accretion makes little contribution to the inflow of a disk gas into the gravitational sphere of a planetary core (see Section~\ref{sec:caseA}), the atmospheric growth of a planet should be limited by the turbulence-driven accretion.
A gas supply to a planet via turbulence-driven accretion, which is smaller than $\sim 10^{-10}~M_\odot~{\rm yr^{-1}}$ at $r \lesssim 1\,{\rm au}$, increases with $r$ (see Figure~\ref{fig:mdot}). As a result, planetary cores within $1\,{\rm au}$ fail to become gas giants and result in super-Earths.
In contrast, they undergo runaway gas accretion to become gas giants in the outer region ($r > 1~{\rm au}$). Thus, the``Beaufort scale'' of wind-driven accretion around a planet influences the habitat of super-Earths and giant planets. 
A mass-period distribution of low-mass planets in the outer region, which will be revealed by future observations (e.g., WFIRST), can test theoretical predictions of our model.

\subsection{Validity of our model}\label{sec:validity}

We discuss the validity of our simplified model. In this study, we considered two cases for the supply of a disk gas to a planetary core; i) accretion flow driven by disk winds can pass through the region around a core (case\,A) and ii) it can fully go into an accreting gas onto the core (case\,B). We found that the accretion of atmospheres can be efficiently limited by the turbulence-drivevn accretion in case\,A. We assumed that planets with mass of 1--10\,$M_\oplus$ are embedded in a disk. We found that the disk scale height ($H$) is larger than a planetary radius, which is defined by the minimum of the Bondi radius and the Hill radius. In calculating the disk scale height, the temperature profile of an optically thin disk in radiative equilibrium, $T = 280 (r/{\rm au})^{-1/2}~{\rm K}$, is used, similar to that in \citet{suzuki_etal16}. Therefore, in this study, super-Earth cores should be sandwiched between disk flows.

As already mentioned in Section~\ref{sec:flow}, we purposely adopt a simplified model for the limit by disk accretion. The model is simple but physically motivated; therefore it is appropriate for the first-step approach. We modeled gas accretion rates onto a core as the minimum of disk accretion and the inflow of a disk gas due to the Kelvin-Helmholtz contraction. The model of disk accretion is explained in Section~\ref{sec:flow}. Here we note that the limit of envelope accretion is more complicated for case\,A. When there is not enough inflow of a disk gas into the Bondi sphere of a planet in case\,A (i.e., $\dot{M}_{\rm env, OMG} < \dot{M}_{\rm disk} = \dot{M}_{\rm turb}$), additional effects should be taken into account in the calculation of the actual gas accretion onto cores. First, as the gas density decreases in the Bondi sphere, the planet may try to gather additional gases from the surrounding disk. As a result, a part of the rapid wind-driven accretion may flow into the accreting gas. Second, there can be additional supply of gas into the lower density region around the planet due to radial viscous diffusion from the inner region, leading to vertical hydrostatic equilibrium. In both cases, the actual envelope accretion rate is higher than $\dot{M}_{\rm turb}$, which corresponds to a case between cases\,A and B. As these effects have not been investigated, three-dimensional hydrodynamic simulations are required to improve our model. Then we also have to include the feedback of the envelope accretion on the disk structure in future work.
 
Our model should also be improved in a different approach in future study. We assume that the envelope accretion rate is given by Eq.~(\ref{eq:mdot_min}) and $\min (\dot{M}_{\rm env, OMG}, \dot{M}_{\rm disk})$. However, it is suggested that only a part of disk accretion into the Bondi sphere can be captured by cores (e.g., \citealt{lubow_dangelo06}; \citealt{tanigawa_tanaka16}). Therefore the disk accretion rate in Eq.~(\ref{eq:mdot_min}) should be multiplied by a reduction factor. In addition, when $\dot{M}_{\rm env, OMG} < \dot{M}_{\rm disk}$, the planet cannot accrete gas to maintain the hydrostatic equilibrium. In this case, the gas in the upper atmosphere may expand, while the envelope contracts due to cooling, leading to an equilibrium state without accreting additional disk gas. In future study, these effects would have to be considered.

\section{Conclusion} \label{sec:conclusions}

We have investigated the formation of close-in super-Earths in a viscously evolving disk with wind-driven accretion by using \textit{N}-body simulations, coupled with gas accretion processes onto a planetary core. 
We considered two cases for the inflow of a disk gas into a planetary core: 
i) only viscous accretion via turbulence (case\,A) and ii) the concurrent accretion in a disk driven by turbulence and wind torque (case\,B). In case\,A, a rapid flow driven by disk winds near the disk surface never gets involved in gas accretion onto a core. A turbulence-driven accretion limits the atmospheric growth of a planet, especially at $r \lesssim 1\,{\rm au}$. 
Even massive cores with $M > 10~M_\oplus$ can be inhibited from undergoing runaway gas accretion. As a result, close-in super-Earths possess atmospheres of $\sim 0.1-10\%$ by mass at the end, which is consistent with bulk compositions of transiting super-Earths inferred from their mass-radius relationships. 
After the depletion of disk gas, the orbital instability among planets results in a non-resonant orbital configuration. Observed distributions (e.g., period ratio) are matched by results of simulations.
On the other hand, in case\,B, a sufficient amount of gas supply from a disk can trigger a rapid gas accretion onto a core. Not super-Earths but multiple gas giants form within $\sim 1\,{\rm au}$, which is incompatible with the fact that hot Jupiters have no siblings.

A limit on gas accretion onto super-Earth cores in case\,A can explain the observed radial distribution of both giant planets and super-Earths. The rate of mass accretion driven by turbulence in a disk increases with radial distance from the star in the disk model based on \citet{suzuki_etal16}. The inflow of a disk gas via turbulence regulates accumulation of a disk gas by planetary cores in the inner region ($r \lesssim 1 {\rm ~au}$), leading to formation of close-in super-Earths rather than hot Jupiters. In contrast, planetary cores are more likely to grow to giant planets in the outer region. Thus, our model has the potential to explain the radial distribution of observed giant planets, which is known to become higher as a radial distance from the star increases.

In this paper, we use a physically motivated but simplified model as a first-step approach to asses effects of disk accretion with disk winds on the envelope growth. With the help of three-dimensional hydrodynamic simulations, the model should be improved in future study.
This paper also contains some caveats for atmospheric retention/loss of a planet during/after the formation stage. We neglected atmospheric erosion processes by giant impacts and stellar XUV irradiations, as stated in Sections~\ref{sec:intro} and \ref{sec:n-body}. In addition, the disk evolution model used in this paper predicts that the surface density of a disk gas in the inner region becomes lower than that for the MMSN model. Depressurization effects caused by a lower gas density of the surrounding disk may enhance mass loss from the upper atmosphere of a planet.
These assumptions make the final atmospheric mass of planets in our simulations an upper limit.
We will consider such a dissipation mechanism in a future work.

\acknowledgments

We would like to thank the anonymous referee for comments that helped us improve the paper.
We would also like to thank Hiroyuki Kurokawa for useful discussions.
This work was supported by JSPS KAKENHI Grant Number 16H07415.

\appendix
\section{Formation of super-Earths in an MRI-inactive disk}\label{sec:app1}

\begin{figure}[ht!]
\plottwo{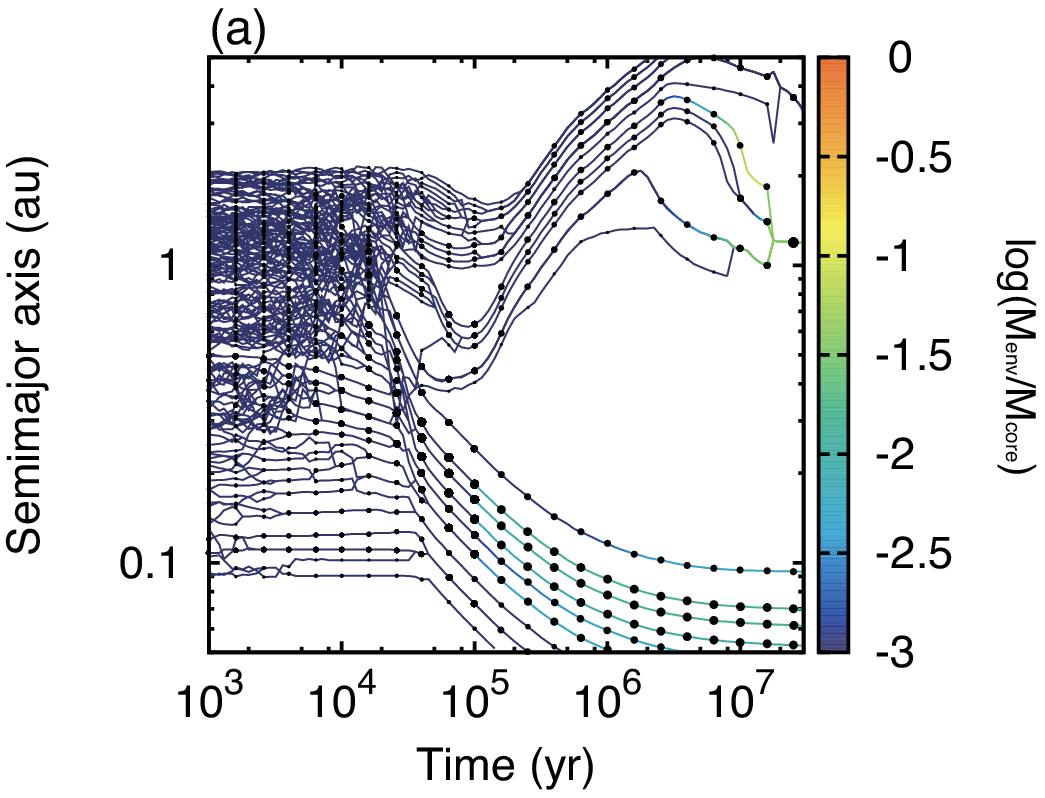}{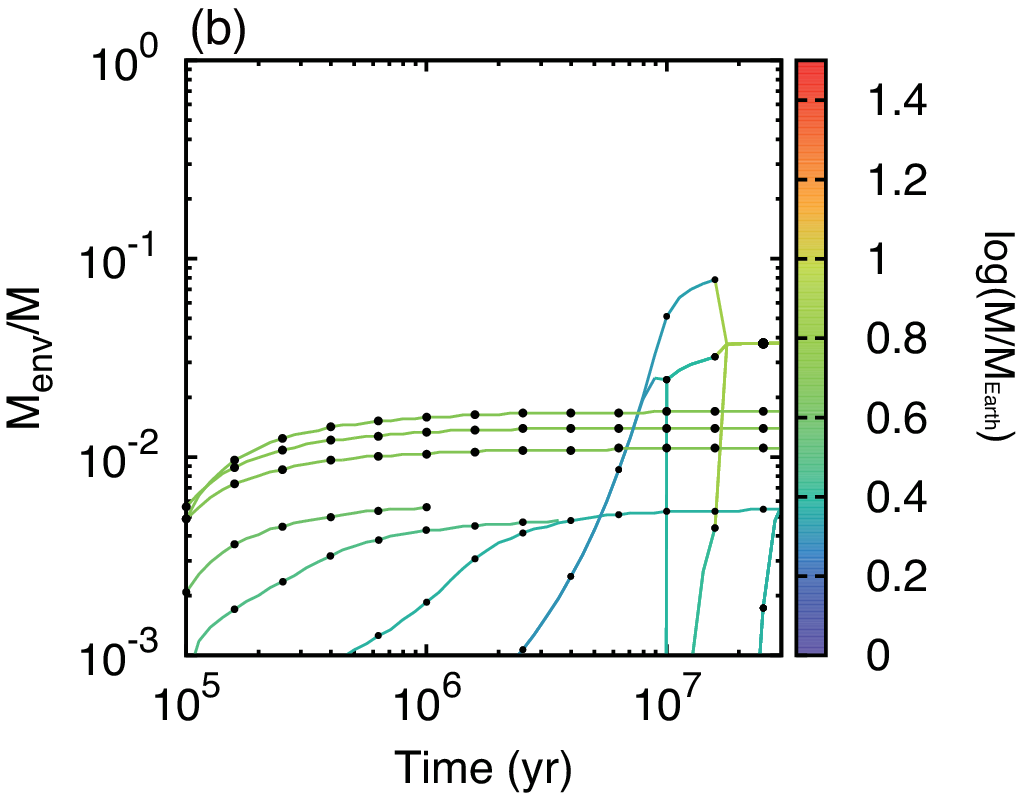}
\caption{Same as Fig.~\ref{fig:run6} but for case\,A and $T_{\rm run}=10^6$ in an MRI-inactive disk.}
\label{fig:run84}
\end{figure}

\begin{figure}[ht!]
\epsscale{0.5}
\plotone{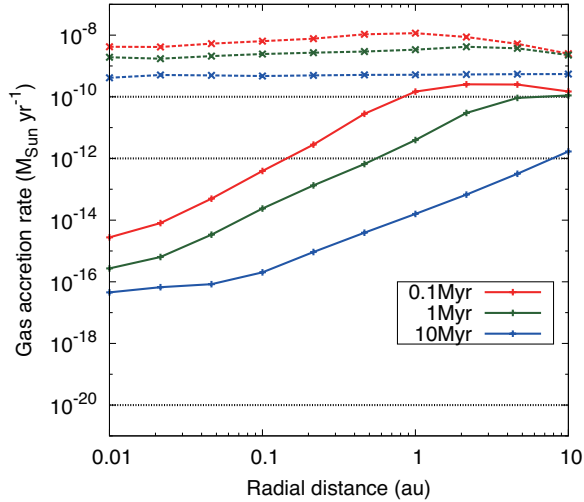}
\caption{Same as Fig.~\ref{fig:mdot} but for MRI-inactive disks.}
\label{fig:mdot_inactive}
\end{figure}

\citet{suzuki_etal16} considered the disk evolution in two cases (i.e., an MRI-active and an MRI-inactive disk). We focus on the late stage of disk evolution and the MRI activity increases with decreasing gas surface density. Therefore, it is reasonable to consider an MRI active disk in this work as discussed in \citet{ogihara_etal18b}.
Here, we also examined formation of super-Earths in an MRI-inactive disk (see also Figure~\ref{fig:r_sigma} for the time evolution of gas surface density). Figure~\ref{fig:run84} shows results of \textit{N}-body simulations for case\,A, where we used $T_{\rm run}=10^6$. 
We find a bimodal distribution of the final locations of planets, which is different from Figure~\ref{fig:run6} in an MRI-active disk. This is because the turbulent viscosity is not large enough to desaturate the corotation torque acted on super-Earths in an MRI-inactive disk,
leading to inward migration of super-Earth mass planets (see \citealt{ogihara_etal18b}).
As well as an MRI-active disk, non of the planets undergo runaway gas accretion for case\,A.
A weaker mass accretion driven by turbulence in an MRI-inactive disk results in the lower mass fractions (0.1--10wt\%) of their H$_2$/He atmospheres, as seen Figure~\ref{fig:mdot_inactive}. This means that an MRI-inactive disk further inhibits formation of planets having massive atmospheres in the close-in region. 

We also perform simulations for case\,B, although figures of these runs are not shown in this paper. As is the case for MRI-active disk, planets can acquire a significant amount of atmospheres for case\,B because mass accretion rates driven by disk winds are almost the same as those in an MRI-active disk (see Figs~\ref{fig:mdot} and \ref{fig:mdot_inactive}).

%Fig.9

%Fig.10

\end{document}